\documentclass[twocolumn,showpacs,preprintnumbers,amsmath,amssymb]{revtex4}

\usepackage{makeidx}
\usepackage{amsmath}
\usepackage{latexsym}
\usepackage{amssymb}
\usepackage{float}

\usepackage{graphicx}
\usepackage{dcolumn}
\usepackage{bm}

\usepackage[dvips]{psfrag}

\psfrag{V}{\hspace{-30 pt}$V\qquad$} \psfrag{S}{$S$} \psfrag{Y}{$Y$}
\psfrag{51}{\hspace{-25 pt}{\scriptsize $5 \times 10^{-8}$}\quad}
\psfrag{101}{\hspace{-21 pt}{\scriptsize $1 \times 10^{-7}$}\quad}
\psfrag{151}{\hspace{-27 pt}{\scriptsize $1.5 \times 10^{-7}$}\quad}
\psfrag{0}{{\scriptsize $0$}\quad}
\psfrag{0.002}{{\scriptsize $0.002$}\quad} \psfrag{0.004}{{\scriptsize $0.004$}\quad}
\psfrag{0.006}{{\scriptsize $0.006$}\quad} \psfrag{0.008}{{\scriptsize $0.008$}\quad}
\psfrag{2}{{\scriptsize $2$}} \psfrag{4}{{\scriptsize $4$}\quad}
\psfrag{6}{{\scriptsize $6$}\quad} \psfrag{8}{{\scriptsize $8$}\quad}
\psfrag{0.5}{{\scriptsize $0.5$}} \psfrag{1}{{\scriptsize $1$}}
\psfrag{1.5}{{\scriptsize $1.5$}} \psfrag{2.5}{{\scriptsize $2.5$}}
\psfrag{3}{{\scriptsize $3$}} \psfrag{10}{{\scriptsize $10$}}
\psfrag{12}{{\scriptsize $12$}} \psfrag{14}{{\scriptsize $14$}}
\psfrag{pot}{\hspace{-18 pt}$10^{10}\times V$}

\psfrag{epsilon}{\hspace{-6 pt}$\epsilon_S, \ |\eta_{SS}|$}
\psfrag{0.2}{{\scriptsize $0.2$}} \psfrag{0.4}{{\scriptsize $0.4$}}
\psfrag{0.6}{{\scriptsize $0.6$}} \psfrag{0.8}{{\scriptsize $0.8$}}
\psfrag{Vt}{\hspace{-6.3 cm}$\displaystyle V(T) \atop {}$}
\psfrag{ReT}{${\rm Re} \ T$} \psfrag{ImT}{${\rm Im}\ T$}
\psfrag{tic1}{ {\scriptsize $5\times 10^{-8}$} } \psfrag{tic2}{ {\scriptsize $1.5\times 10^{-7}$} }
\psfrag{tic3}{ {\scriptsize $2.5\times 10^{-7}$} }  
\psfrag{Vtr}{ \hspace{-0.7 cm}$10^{-7}\times V({\rm Re} \ T) $}
\psfrag{tic4}{ {\scriptsize \hspace{0.2 cm}$1.0$} } \psfrag{tic5}{ {\scriptsize \hspace{0.2 cm}$3.0$} }

\psfrag{q}{$\!{}_{{}_2}$}

\begin{document}

\preprint{TOK-HEP051220}

\title{Angular Power Spectrum and Dilatonic Inflation 
\\in Modular-Invariant Supergravity}

\author{Mitsuo J. Hayashi}
 \email{mhayashi@keyaki.cc.u-tokai.ac.jp}
\affiliation{
Department of Physics, Tokai University,\\
\mbox{1117~Kitakaname,~Hiratsuka, Kanagawa,~259-1292, Japan}
}

\author{Shiro Hirai}
\email{hirai@isc.osakac.ac.jp}
\affiliation{%
Department of Digital Games,
Osaka Electro-Communication University,\\
1130-70 Kiyotaki, Shijonawate, Osaka, 575-0063, Japan
}%

\author{Yusuke Okame}%
 \email{5atrd001@keyaki.cc.u-tokai.ac.jp}
\affiliation{%
Graduate School of Science and Technology, Tokai University,\\
1117 Kitakaname, Hiratsuka, Kanagawa, 259-1292, Japan
}%

\author{Tomoki Watanabe}%
 \email{tomoki@gravity.phys.waseda.ac.jp}
\affiliation{%
Advanced Research Institute for Science and Engineering, Waseda University,\\
3-4-1 Okubo, Shinjuku-ku, Tokyo, 169-8555, Japan
}%

\date{\today}

\begin{abstract}
The angular power spectrum is investigated in the model of supergravity, incorporating the target-space duality 
and the non-perturbative gaugino condensation in the hidden sector.
The inflation and supersymmetry breaking occur at once by the interplay between the dilaton field as inflaton and the condensate
gauge-singlet field. The model satisfies the slow-roll condition which solves the $\eta$-problem.
When the particle rolls down along the minimized trajectory of the potential  at a duality invariant fixed point $T=1$,
we can obtain the $e$-fold value $\sim 57$.
And then the cosmological parameters obtained from our model well match with the recent WMAP data combined with other experiments.
The $TT$ and $TE$ angular power spectra  also show that
our model is compatible with the data for $l > 20$.
However,
the best fit value of $\tau$ in our model is smaller than that of the $\Lambda$CDM model. 
These results suggest that, among supergravity models of inflation, the modular-invariant supergravity
seems to open a hope to construct the realistic theory of particles and cosmology.
\end{abstract}

\pacs{04.65.+e, 11.25.Mj, 11.30.Pb, 12.60.Jv, 98.80.Cq}
\maketitle

\section{\label{sec:level1}Inflationary Cosmology}

Since WMAP combined with the other experiments demonstrated on February, 2003 that the big bang and
inflation theories continue to be true\cite{ref:1}, 
the constraints on the cosmological parameters, such as the spectral index and
its running as well as the ratio of the tensor to the scalar, have been improved by combining WMAP
and Ly$\alpha$ forest\cite{ref:2,ref:3}. 
Recently the polarization-temperature angular cross power spectrum of the cosmic microwave
background (CMB) from the 2003 Flight of Boomerang was also published\cite{ref:4}.

From the theoretical viewpoint,
it is customary to introduce scalar field(s) called inflaton
into inflation models\cite{ref:5,ref:6};
there are, however, several problems in constructing successful theories:
i) What is it, the inflaton?
ii) What kind of theoretical frameworks is the most appropriate as the theory of particle physics,
inflation and the recently observed accelerating universe?
iii) How to explain the contents of the universe?: 
Baryonic matter 4\%, Dark matter 23\%, Dark energy 73\% and so on.
These problems seem to require far richer structures of contents than those of the standard
theory of particles.
And more phenomenologically,
iv) Is the model consistent with the observed CMB angular power spectra?
In particular, the inflaton should satisfy the slow-roll condition in order that the model
predict the nearly scale-invariant spectral index as well as the sufficient number of e-folds.
(See ref.\cite{ref:5} for the recent review on the theories of inflation.)

Although recently the inflation theories in string theories are extensively and promissingly investigated  by various authors\cite{ref:5},
 here we concentrate on the framework of a supergravity inspired by superstrings, following previous papers\cite{ref:7,ref:8,ref:9}. 
The well-known difficulty of supergravity is that the potential form gives arise the $\eta$-problem\cite{ref:8},
 which breaks the slow-roll condition. The string-inspired supergravity is
derived from the $d=10$ 
heterotic string by dimensional reduction to $N=1,\ d = 4$ supergravity\cite{ref:7}, 
whose typical features are: i) No-scale structure at the tree level. ii) $E_8\times E_8$ gauge group (one of the $E_8$ is called
 the hidden sector of the gauge group). iii) Non-perturbative gaugino condensation in the hidden sector can break the supersymmetry. 
iv) Modular invariance, acting on a single modulus  $T$, valid at any string-loop order (Target-space duality)\cite{ref:8,ref:9,ref:10,ref:11,ref:12,ref:13}.
   
In this paper, the angular power spectrum in modular-invariant supergravity is investigated in the model
where we had shown that the inflation and the supersymmetry breaking occur at once
by the interplay between the dilaton field as the inflaton and the condensated gauge-singlet field rolling down the inflationary trajectory,
free from the $\eta$-problem\cite{ref:9}.

\section{A String-inspired Supergravity}

First of all, for the self-containedness, we will review the idea of the construction of the effective theory of gaugino condensation,
incorporating the target-space duality, following ref.\cite{ref:11},
where the gaugino condensation has been described by a duality-invariant effective action
for the gauge-singlet gaugino bound states coupled to the fundamental fields as the dilaton $S$ and moduli $T$.
On the other hand, in ref.\cite{ref:10}, the gaugino-condensate has been replaced by its vacuum expectation value
to yield a duality-invariant ``truncated" action that depends on the fundamental fields only.
The equivalence between these two approaches had been proved in refs.\cite{ref:12}.

 Assuming that the compactification of the superstring theory preserves $N=1$ supersymmetry, the effective theory should be of the general type of $N=1$ supergravity coupled to gauge and matter fields. 
The most general form of Lagrangian in $N=1$ supergravity at the tree-level is\cite{ref:7}(See also \cite{ref:8,ref:14,ref:15,ref:16}):
\begin{equation}
\mathcal{L}=
-\frac{1}{2}\left[e^{-K/3}S_0\bar{S}_0\right]_D
+\left[S_0^3W\right]_F
+\left[f_{ab}W^a_\alpha \epsilon^{\alpha\beta}W^b_\beta\right]_F,
\end{equation}
where the K\"{a}hler potential $K$ is given by
\begin{equation}
K=-\ln \left(S+S^\ast\right)-3\ln \left(T+T^\ast-|\Phi_i|^2\right),
\end{equation}
and the gauge function $f_{ab}$ is
\begin{equation}
f_{ab}=\delta_{ab} S.
\end{equation}
In order to construct the effective theory of gaugino condensation, we introduce the composite superfield $Y$
 of the gaugino condensation\cite{ref:11,ref:13}:
\begin{equation}
Y^3=
\delta_{ab}W^a_\alpha \epsilon^{\alpha\beta}W^b_\beta/S_0^3
=(\lambda\lambda+\cdots)/S_0^3,
\end{equation}
where $\lambda$ is the gaugino fields in the Hidden sector.

The effective K\"{a}hler potential and superpotential incorporating modular invariant one-loop corrections are given as\cite{ref:11}:
\begin{equation}
K=-\ln \!\left(S+S^\ast\right)
-3\ln \!\left(T+T^\ast-|Y|^2-|\Phi_i|^2\right),
\end{equation}
and
\begin{equation}
W=3bY^3\ln\left[c\>e^{S/3b}\>Y\eta^2(T)\right]+W_{\rm matter}
\end{equation}
where $\eta$ is Dedekind's $\eta$ function, $c$ is a free parameter in the theory and $b=\frac{\beta_0}{96\pi^2}$
($\beta_0$ is the one-loop beta-function coeffiients).

Since $\langle S+S^\ast\rangle =\alpha^\prime m_{\rm pl}^2$, the choice:
\begin{equation}
[e^{-K/3}S_0\bar{S}_0]_{\theta=\bar\theta=0}=[S+\bar{S}]_{\theta=\bar\theta=0},
\end{equation}
is corresponded to the conventional normalization of the gravitational action:
\begin{equation}
\mathcal{L}_{\rm grav} \sim [e^{-K/3}S_0\bar{S}_0]_{\theta=\bar\theta=0}R.
\end{equation}

Then, the scalar potential is obtained as follows:
\begin{widetext}
\begin{eqnarray}
V(S,T,Y)
&=&
\frac{3(S+S^\ast)|Y|^4}{(T+T^\ast-|Y|^2)^2}
\Bigg(
	3b^2 \left| 1 + 3 \ln \left[ c \>e^{S/3b}\> Y \eta^2(T) \right] \right|^2
\nonumber\\
&&\hspace{-1.9cm}
	+ \frac{|Y|^2}{T+T^\ast-|Y|^2}
	\Bigg|
		S + S^\ast - 3b \ln \left[ c\>e^{S/3b}\>Y\eta^2(T) \right]
	\Bigg|^2
	+ 6b^2 |Y|^2
	\Bigg[
		2(T + T^\ast) \left| \frac{\eta^\prime (T)}{\eta(T)} \right|^2
		+ \frac{\eta^\prime (T)}{\eta(T)}
		+ \left( \frac{\eta^\prime (T)}{\eta(T)} \right)^\ast
	\Bigg]
\Bigg),
\end{eqnarray}
\end{widetext}
where matter fields are neglected.

\section{Inflationary Trajectory and Stability in modular variable $T$}

The potential is modular invariant and shown to be stationary at the self-dual points $T=1$ and $T=e^{i\pi /6}$. We found that the potential $V(S,Y)$ at $T=1$ has a stable minimum at $(Y_{\rm min},S_{\rm min})\sim (0.00646, 0.435)$ (See Fig.1).

Therefore, we may conclude that
inflation arises
by the evolution of dilaton field $S$
and supersymmetry is broken
by the condensated field $Y$,
provided it begins at the unstable saddle point and slowly rolls down to the minimum.

\begin{figure}[h]
\begin{center}
\includegraphics[scale=0.77]{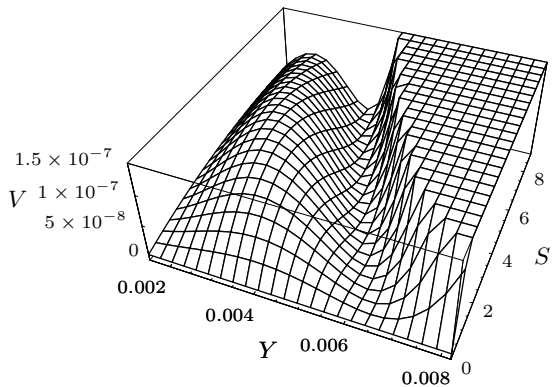}
\end{center}
\caption{The plot of $V(S,Y)$ at fixed $T=1$ (self-dual point) with $c=183,\ b=5.5$.The stable minimum of $V_Y(S)=0$ and a saddle point exist.
We can see a valley of the potential and a stable minimum of  
$V_Y(S)=0$ at $(Y_{\rm min},S_{\rm min})\sim (0.00646, 0.435)$. 
}
\end{figure}

The inflationary trajectory will be well approximated by the equation:
\begin{equation}
Y_{\rm min}(S)\sim 0.00663e^{-S/16.2}.
\end{equation}

In Fig. 2, we have shown a plot of $V(S)$ minimized with respect to $Y$.
As shown by Ferrara {\it et al.}\cite{ref:11}, supersymmmetry is broken by the hidden sector gaugino condensation\cite{ref:13} because 
$\langle |F| \rangle \propto \langle |\lambda\lambda |\rangle\neq 0$. 

\begin{figure}[H]
\begin{center}
\includegraphics[scale=0.8]{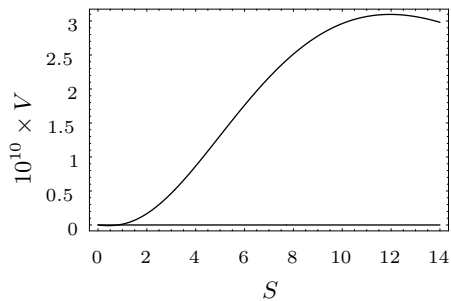}
\end{center}
\caption{The plot of $V(S)$ minimized with respect to $Y$. The minimum value of the potential is $V(S_{\rm min})\sim -9.3\times 10^{-13}$.}
\end{figure}

One of the main purposes of this paper is to prove that the dilaton field plays the role of the inflaton field.

The slow-roll parameters (in Planck units $m_{\rm pl}/\sqrt{8\pi}=1$) are defined by:
\begin{equation}
\epsilon_\alpha=\frac{1}{2}\left(\frac{\partial_\alpha V}{V}\right)^2
\quad ,\quad 
\eta_{\alpha\beta}=\frac{\partial_\alpha\partial_\beta V}{V}.
\end{equation}
The slow-roll condition demands both values to be lower than 1. It is the end of inflation,
when the slow-roll parameter $\epsilon_\alpha$ reaches the value 1. After passing through the end of inflation,
``matter" may be produced during the oscillations around the minimum of the potential (reheating) with the critical density,
i.e. $\Omega=1$.
Although any successful theory of inflation should explain the mechanism of
the reheating process, we remain this reheating problem for later work in the
framework of the present model.

The values of $\epsilon_S$ and $\eta_{SS}$ are obtained numerically in Fig. 3 fixing the parameters $c=183\ {\rm and}\ b=5.5$;
we find the slow-roll condition is well satisfied, and the $\eta$-problem can just be avoided.

\begin{figure}[H]
\begin{center}
\includegraphics[scale=0.8]{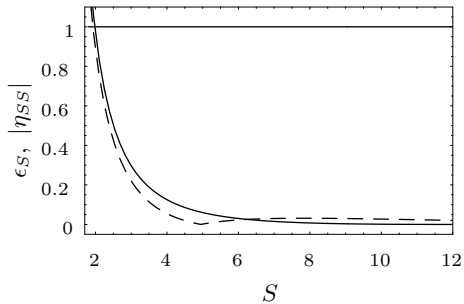}
\caption{The evolution of the slow-roll parameters.
The solid curve represents $\epsilon_S$ whereas the dashed curve denotes $|\eta_{SS}|$, which
demonstrate that the potential $V(S)$ is sufficiently flat.
Inflation ends at $S \sim 1.98$ in our model.}
\end{center}
\end{figure}

%

The potential $V$  is stable at the self-dual point $T=1$ in arbitrary points in the inflationary trajectory for our choice of the parameters $c$ and $b$.
By choosing the three points, i.e., horizon exit,  end of inflation and the stable minimum and inserting those values $S,\ Y$ at these points into the original $V(S,Y,T)$, we will here demonstrate that the potential $V(T)$ has minima exactly  at $T=1$ and hence is stable at these typical stages in the inflationary trajectory.
 The variations of $V(T)$ are obtained numerically in Figs. 4 and 5 at the fixed parameters $c=183\ {\rm and}\ b=5.5$.
 
\begin{figure}[H]
\begin{center}
\includegraphics[scale=0.8]{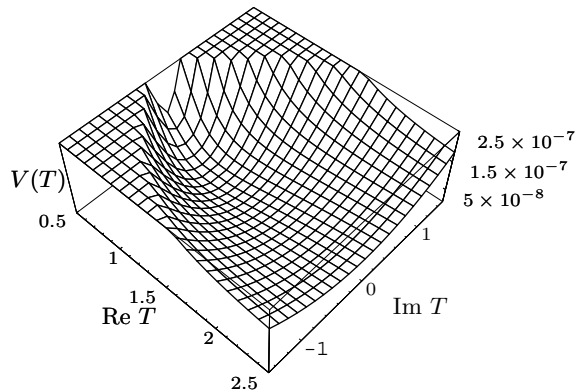}
\caption{The 3D plot around the minimum of $V(T)$ as a function of complex variable  $T$ for $S_{\rm min}$ and $Y_{\rm min}$.
${\rm Im}\ T=0$ is obviously stable.}
\end{center}
\end{figure}

\begin{figure}[H]
\begin{center}
\includegraphics[scale=0.8]{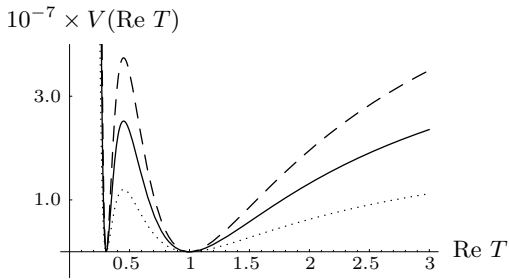}
\caption{The plots of $V(T)$ at ${\rm Im}\ T=0$ at three representative inflationary stages. It is obvious that $T=1$ is stable.
The solid, dashed and dotted curves represent the stages at the horizon exit, the end of inflation and the stable minimum respectively. }
\end{center}
\end{figure}

Number of $e$-folds at which a comoving scale $k$ crosses the Hubble scale $aH$ during inflation is given by:
\begin{equation}
N(k)\sim
62 - \ln\frac{k}{a_0H_0} 
-\frac{1}{4}\ln\frac{(10^{16}\>{\rm GeV})^4}{V_k}
+\frac{1}{4}\ln\frac{V_k}{V_{\rm end}},
\end{equation}
where we assume $V_{\rm end}=\rho_{\rm reh}$. We focus on the scale $k_*=0.05\> {\rm Mpc^{-1}}$ and the inflationary energy scale is 
$V\sim10^{-10}\sim(10^{16}\>{\rm GeV})^4$ 
as shown in Fig. 2, therefore the number of $e$-folds which corresponds to our scale must be around 57. 

On the other hand, using the slow-roll approximation (SRA), $N$ is also calculated by:
\begin{equation}
N\sim -\int^{S_2}_{S_1}\frac{V}{\partial V}dS.
\end{equation}
We could have obtained the number $\sim 57$, by integrating from  
$S_{\rm end}\sim 1.98$ to $S_*\sim 10.46$,
 fixing the parameters $c=183\ {\rm and}\ b=5.5$,  
 i.e. our potential has the ability to produce the cosmologically plausible number of $e$-folds. Here $S_*$ is the value corresponding to $k_*$.

Next, the scalar spectral index standing for a scale dependence of the spectrum of density perturbation and its running are defined by:
\begin{eqnarray}
n_s-1&=&\frac{d\ln \mathcal{P_R}}{d\ln k}, \\
\alpha_s&=&\frac{dn_s}{d\ln k}.
\end{eqnarray}
These are approximated in the slow-roll paradigm as:
\begin{eqnarray}
n_s(S)&\sim& 1-6\epsilon_S+2\eta_{SS}, \\
\alpha_s(S)&\sim& 16\epsilon_S\eta_{SS} -24\epsilon_S^2-2\xi^2_{(3)},
\end{eqnarray}
where $\xi_{(3)}$ is an extra slow-roll parameter that includes trivial third derivative of the potential.  
Substituting $S_*$ into these equations, we have $n_{s*}\sim0.95$ and $\alpha_{s*}\sim -4\times10^{-4}$. 

Because $n_s$ is not equal to 1 and $\alpha_{s}$ is almost negligible, our model supports the model with tilted power law spectrum. The value of $n_{s*}$ is consistent with the recent observations; 
 the best fitting of them (WMAPext, 2dFGRS and Lyman $\alpha$) for power law $\Lambda$CDM model suggests \cite{ref:1,ref:2,ref:3,ref:4} 
  $n_s(k_*)=0.96\pm0.02$.

Finally, estimating the spectrum of the density perturbation caused by slow-rolling dilaton\cite{ref:19}:
\begin{equation}
\mathcal{P_R}\sim\frac{1}{12\pi^2}\frac{V^3}{\partial V^2},
\end{equation}
we find $\mathcal{P_R}_*\sim2.1\times10^{-9}$. 

This result matches the measurements as well. Incidentally speaking, the energy scale $V\sim10^{-10}$ is also within the constrained range obtained by Liddle and Leach\cite{ref:17}.\\

Gravitational waves are inevitable consequence of all inflational models.
Now the tensor purturbation and the gravitational wave spectrum is given as:
\begin{equation}
\mathcal{P}_{\rm grav}=8\left(\frac{H}{2\pi}\right)^2
=\frac{2}{3\pi^2}V.
\end{equation}
In SRA, the spectral index of $\mathcal{P}_{\rm grav}$ is given by the slow-roll parameters $\epsilon$ and $\eta$ as:
\begin{equation}
n_{T}=-2\epsilon
\label{eq:T18}.
\end{equation}
The ratio $r$ between $\mathcal{P}_{\mathcal{R}}$ and $\mathcal{P}_{\rm grav}$ is given as
\begin{equation}
r=\frac{\mathcal{P}_{\rm grav}}{\mathcal{P}_{\mathcal{R}}}=16\epsilon
=-8n_{T}.
\end{equation}
The gravitational wave spectrum does not evolve and remains frozen-in as a massless field even after the horizon-exit, independent of the scalar perturbations\cite{ref:18}. Contrary to this fact, the primordial curvature fluctuation $\mathcal{R}$ evolution is given by the product between the transfer function $T_r(k)$ and $\mathcal{R}$ :
\begin{equation}
\mathcal{R}_{\scriptsize\textbf{k}}^{(m)}=T_r(k) \mathcal{R}
_{\scriptsize\textbf{k}}.
\end{equation}
Therefore, the ratio $r$ evolves as
\begin{equation}
\left(\frac{\mathcal{P}_{\rm grav}}{\mathcal{P}_{\mathcal{R}}}\right)^{(m)}
=-8T_r{}^2n_{T}
\end{equation}
up to the present time.
This result will be used in the calculation of the angular power spectra.

\section{The Angular Power Spectrum of the model}

In our model we can calculate the angular power spectrum to compare with the WMAP analysis
and the other experimental data\cite{ref:1,ref:2,ref:3,ref:4,hira}.
The multipoles $a_{lm}$ of the CMB anisotropy are defined by
\begin{eqnarray}
\Delta T &\equiv&
\frac{\delta T}{T}=\sum_{l>0}\sum_{m=-l}^{m=l}a_{lm}Y_{lm}(\bf{e}),
\\
a_{lm}&=&\int d\Omega_{\bf{n}} \Delta T( {\bf n}) Y_{lm}^*(\bf{e}),
\end{eqnarray}
where $Y_{lm}(\bf{n})$ are the sperical harmonic functions evaluated in the direction $\bf{n}$. The multipoles with $l\geq 2$ represent the intrinsic anisotropy of the CMB.
If the CMB temperature fluctuation $\Delta T$ is Gaussian distributed, then each $a_{lm}$ is an independent Gaussian deviate with
\begin{equation}
\langle a_{lm} \rangle =0,
\end{equation}
and
\begin{equation}
\langle a_{lm}a_{l'm'}^* \rangle =\delta_{ll'}\delta_{mm'}C_l,
\end{equation}
where
$C_l$ is the ensemble average power spectrum, or, the angular power spectrum of the CMB.
In general, the cosmological information is encoded in the standard deviations and correlations
of the coefficients:
\begin{equation}
\left \langle X Y \right \rangle = \left \langle a_{lm}^X a_{l'm'}^Y {}^* \right \rangle
= \delta_{ll'} \delta_{mm'} C_{l}^{XY}.
\end{equation}

If we use the spherical expansion of the form for an arbitrary function $g(\bf{x})$:
\begin{equation}
g({\bf x})=\int_0^\infty dk\sum_{lm}g_{lm}(k)\sqrt{\frac{2}{\pi}}kj_l(kx)Y_{lm}(\theta,\phi),
\end{equation}
where $j_l$ is the spherical Bessel function,$(\theta,\phi)$ is the direction of ${\bf x}$,
then the angular power spectrum $C_l^{TT}$ and the temperature-polarization  cross power spectrum $C_l^{TE}$
will be given by
\begin{eqnarray}
C_l^{TT} &=& 4\pi \int_0^\infty T_\Theta^2(k,l)\mathcal{P}_{\mathcal{R}}(k)\frac{dk}{k},
\\
C_l^{TE} &=& 4 \pi \int_0^\infty T_\Theta (k,l) T_E (k,l) \mathcal{P}_{\mathcal{R}} (k) \frac{dk}{k},
\end{eqnarray}
where $T_\Theta$ and $T_E$ are the transfer functions and $\Theta$ is the brightness function.


Now we will show the behavior of those power spectra in our model.

The scalar spectral index is $n_s(k_*)=0.95$ and the running index is $\alpha_s(k_*)=-0.0004$
at $k_*=0.05 \ {\rm Mpc^{-1}}$ as already shown.
The tensor-to-scalar ratio is assumed as
$r=16\epsilon=8n_T=0.00923$ ($\epsilon=0.00058$ at $k_*=0.05 \ {\rm Mpc^{-1}}$).

We will use the CMBFAST\cite{cmbfast}, where we have  assumed the cosmological parameters to be:
$\Omega_{\rm tot}=1$ for the total enegy density, 
$\omega_\Lambda =-1$ and $\Omega_\Lambda=0.73$ for the dark energy,
$\Omega_{\rm b}=0.046$ and $\Omega_{\rm cdm}=0.224$ for the baryonic and dark matter density,
$h=0.71$ for the Hubble constant. 
The angular $TT$ power spectra were normalized with respect to 11 data points in the WMAP
data from $l=65$ to $l=210$ 
and
the same values are used in the analysis of the angular $TE$ spectrum.

By using the likelihood method\cite{likelihood}, we calculated the $\chi^2$ values for the $TT$ and $TE$ spectrum
and their total sum, which are shown at Table I.

The ${\Lambda}$CDM model shows
that the best fit value of $\tau$ is 0.17 by the same method.
On the other hand, the best fit of our model seems realized at $\tau= 0.13$ for $TE$ mode, which is allowable within the experimental error,
while the total $\chi^2$ value takes minimum at $\tau=0.07$.

The angular power spectra of our model are presented in Fig. 6 for $TT$ mode, Fig. 7 for $TE$ mode at the values $\tau= 0.17, \ 0.07$
and Fig.8 for $TE$ mode at the values $\tau= 0.17, \ 0.13, \ 0.07$ for $l\leq 20$ with more detailed data.
Because the $TE$ spectra almost completely coincide with those of the $\Lambda$CDM model for $20 < l < 500$,
we have shown the spectrum for $l < 50$ in the Figs.7 and 8,
where a distinction between our model and the $\Lambda$CDM model can be seen.
Moreover, the best fit value of $\tau$ takes 0.13 for $TE$ mode in our model whereas 0.17 in the $\Lambda$CDM model.
We would like to emphasize this result as one of the characteristic features in our model. 

For $TT$ mode, although both our model and the $\Lambda$CDM model can explain almost satisfactorily the WMAP data,
there remains some inconsistency in the suppression of the spectrum at large angular scales ($l=2,3$)\cite{hira}.
These data points appear to be the reason why the best fit of $\tau$ even becomes 0 for $TT$ mode and 0.07 for the total $\chi^2$ value,
lower than 0.13 which is the best $\chi^2$ value for $TE$ mode. 

In summary,
the model we have here investigated is compatible with the present observational data
for $l > 20$, whereas there remains some problems unexplained for small $l$.
 
\begin{table}[h]
	\begin{center}
		\caption{The $\chi^2$ values for the $TT$ and $TE$ spectrum and their total sum.
		The best fit is at $\tau= 0.13.$ for $TE$ mode and at $\tau=0.07$ for the total sum.
		\vspace{0.5mm}}
		\begin{tabular}{c|c|c|c}
\hline
$\tau$ & TT & TE & Total \\
\hline
\hline
 {\ } 0.17 {\ }  & {\ } 986.92 {\ } & {\ } 456.73 {\ } & {\ } 1443.65 {\ } \\
 0.14 & 982.10 & 456.50 & 1438.60 \\
 0.13 & 980.87 & 456.48 & 1437.35 \\
 0.12 & 980.00 & 456.72 & 1436.71 \\
 0.08 & 977.87 & 458.28 & 1436.15 \\
 0.07 & 977.40 & 458.63 & 1436.03 \\
 0.06 & 977.31 & 459.51 & 1436.81 \\
 0.01 & 976.47 & 464.53 & 1441.00 \\
 0.00 & 976.26 & 465.35 & 1441.61 \\
\hline
%
%
%
%
%
%
		\end{tabular}
	\end{center}
\end{table}
\vspace{-0.7 cm}
\begin{figure}[H]
		\begin{center}
		\rotatebox[origin=c]{-90}{
		\includegraphics[trim=2cm 1cm 2.5cm 3.5cm, clip, width=.645\linewidth]{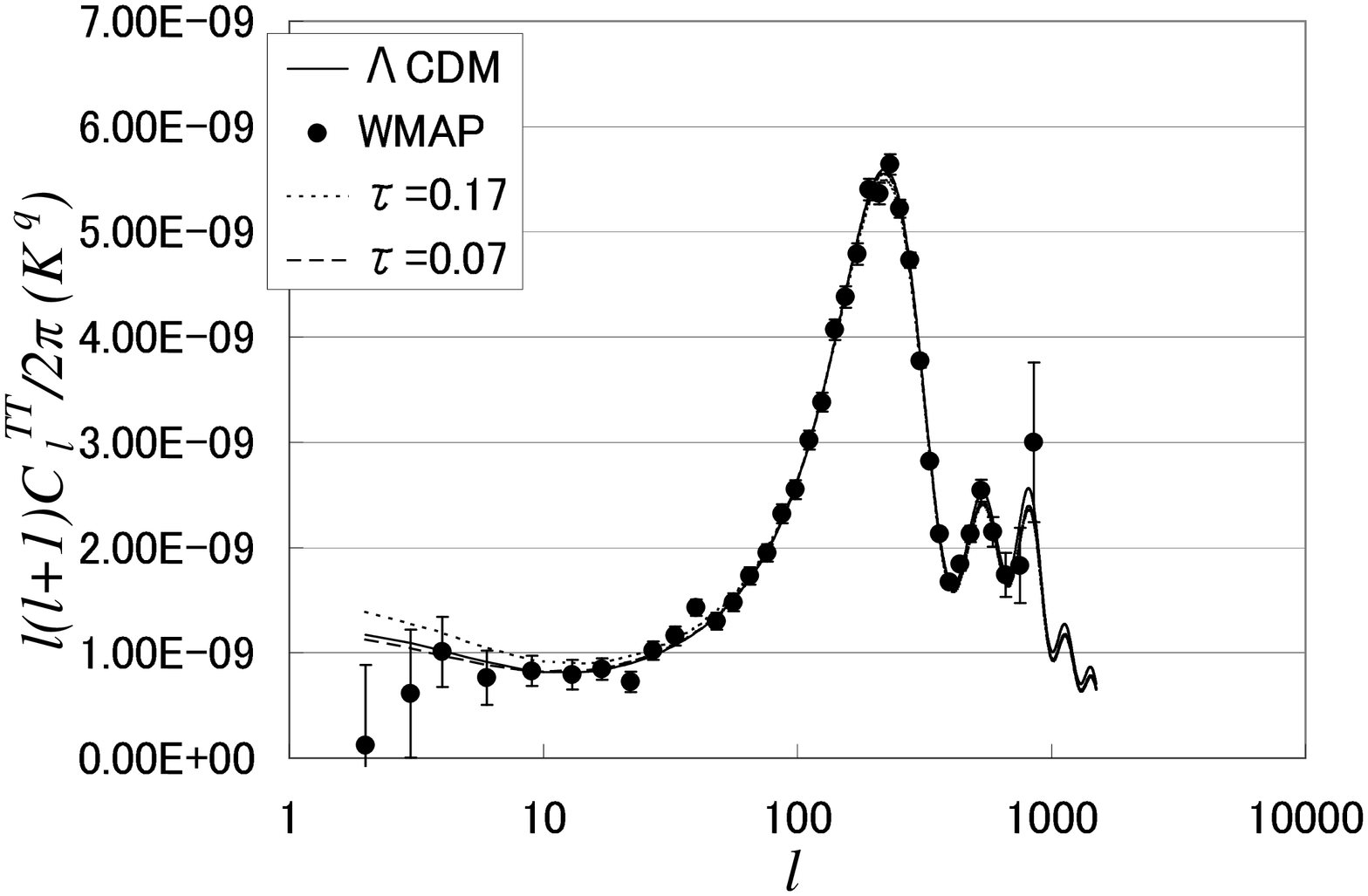}
		}
		\vspace{-2 cm}
		\caption{Temperature angular power spectrum ($TT$).}
		\end{center}
\end{figure}
\vspace{-1 cm}
\begin{figure}[h]
	\begin{center}
	\rotatebox[origin=c]{-90}{
	\includegraphics[trim=1cm 1cm 2.5cm 3.5cm, clip, width=.68\linewidth]{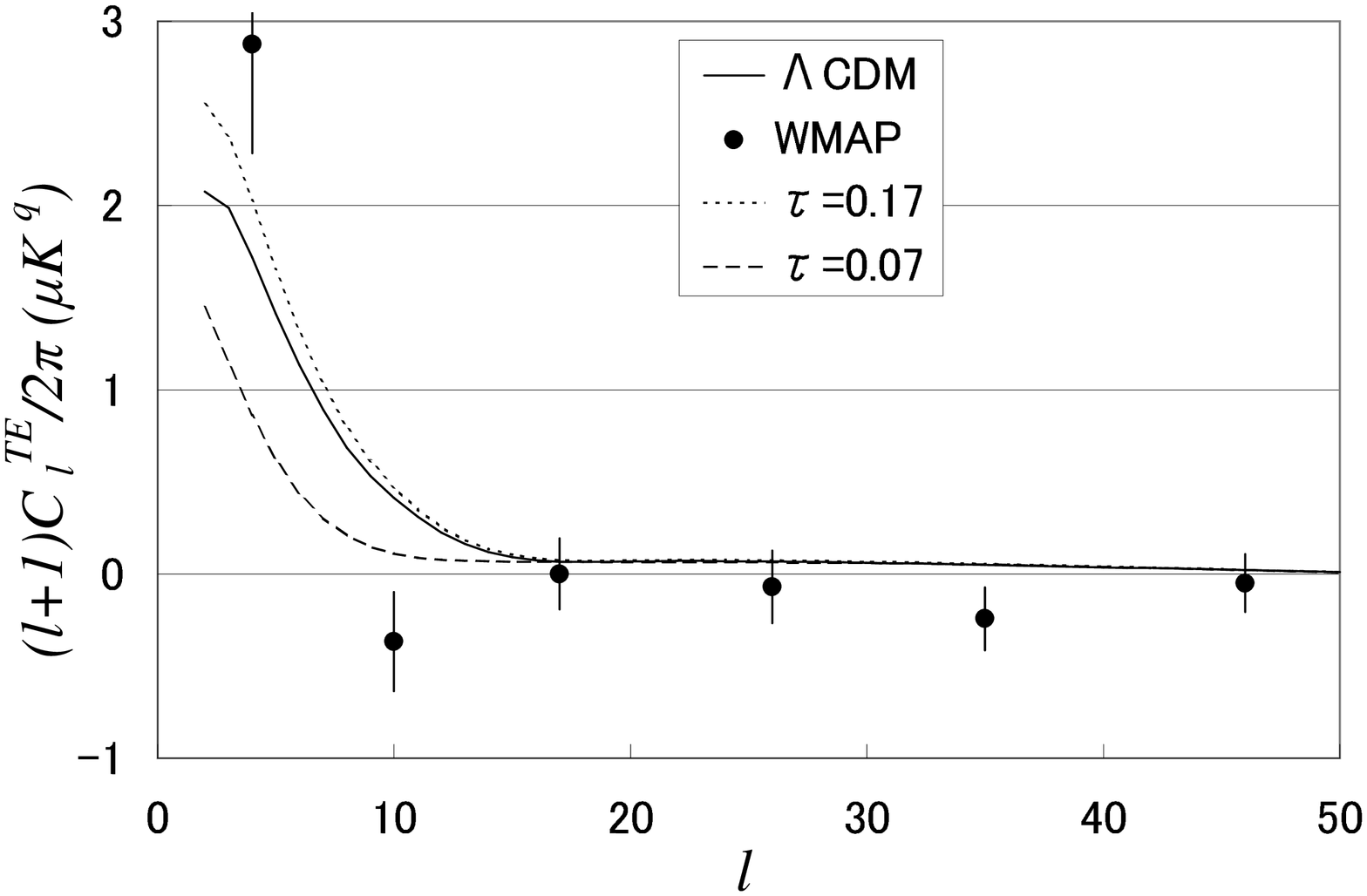}
	}
	\vspace{-1.5 cm}
	\caption{Temperature-polarization cross power spectrum ($TE$) for $l < 50$.
	}
	\end{center}
\end{figure}
\begin{figure}[H]
	\begin{center}
	\rotatebox[origin=c]{-90}{
	\includegraphics[trim=1cm 1cm 2.5cm 3.5cm, clip, width=.68\linewidth]{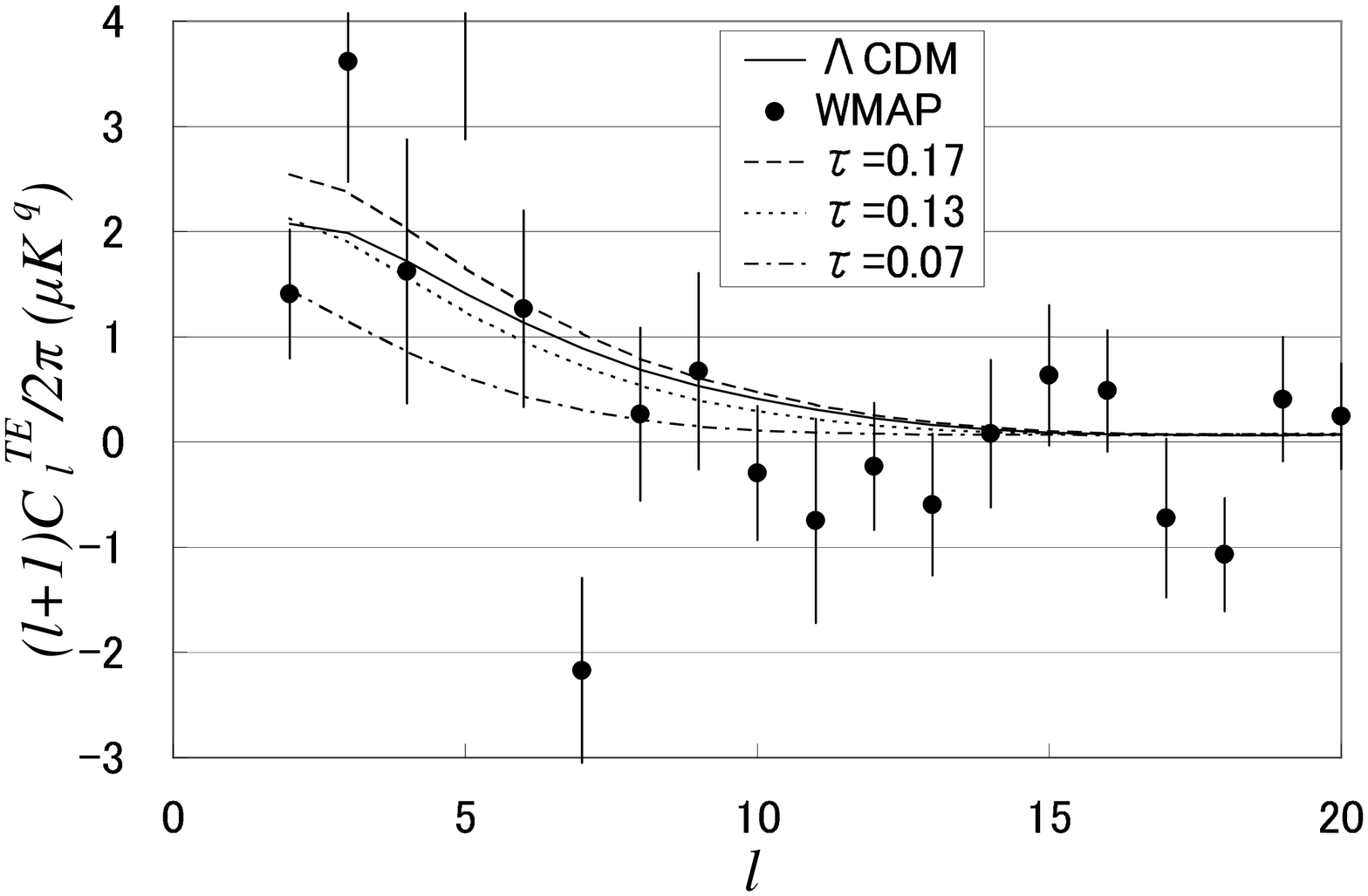}
	}
	\vspace{-1.5 cm}
	\caption{Temperature-polarization cross power spectrum ($TE$) for $l < 20$ with more detailed data.
	For larger values of $l$ our model almost completely coincide with the $\Lambda$CDM model.
	}
	\end{center}
\end{figure}

\section{Conclusions}

 Now we conclude that inflation and supersymmetry breaking occur at once by the interplay between the dilaton field as the inflaton
and the condensate gauge-singlet field. 
Our model is compatible with the angular power spectra of the WMAP data
for $l > 20$, whereas there remains some problems unexplained for small $l$, as the $\Lambda$CDM model is.
The best fit value of $\tau$ in our model is smaller than 0.17 in the $\Lambda$CDM model.

It appears that supergravity is one of the most plausible frameworks to explain the new physics,
including the undetected objects, such as the inflaton, dark matter and dark energy. 
Particularly, since the inflaton field is concerned with the Planck scale physics,
the dilaton field seems to be the most presumable candidate of the inflaton.
Among the possibe supergravity models of inflation, the modular invariant model
here revisited seems to open a hope to construct the realistic theory of particles and cosmology.

For further investigations,
i) We should consider on the effects of the hidden sector massive matter
over inflation and the supersymmetry breaking\cite{ref:12,ref:13}.
ii) It will be interesting to understand what kind of phenomena are obtained
from the S-duality invariant theory\cite{ref:15}.
Furthermore,
iii) What kind of theories can shed light on the great problem of the dark energy and dark matter to understand their origin
and their relation to the recently observed accelrating universe
\cite{ref:22, ref:23} and \cite{ref:1,ref:2,ref:3,ref:4}? 
iv) Gravitino, inflatino and axion production and their effects should be
traced\cite{ref:21}.
v) Brane world cosmology and M-theoretical approach 
might be promising.
(See Linde's lecture in ref.\cite{ref:5}, ref.\cite{ref:15,ref:20} and references therein.)
These problems and the reheating after inflation will be our further tasks\cite{ref:5}.

\vspace{12 pt}
\begin{acknowledgments}
We are grateful to Prof. T. Takami of Osaka Electro-Communication University for his help on numerical calculations.
\end{acknowledgments}

\end{document}